\documentclass[aps,prd,onecolumn,groupedaddress,showpacs,nofootinbib,amssymb]{revtex4}
\usepackage[dvips]{graphicx}
\usepackage{amssymb}
\usepackage{amsmath}
\usepackage{graphicx,,color}
\usepackage{amsfonts}
\usepackage{bm}
\usepackage{cancel}
\usepackage{comment}
\newcommand\be{\begin{equation}}
\newcommand\ee{\end{equation}}

\allowdisplaybreaks[4]

\begin{document}

\title{Reconstruction of Slow-roll $F(R)$ Gravity Inflation from the Observational Indices}
\author{S.~D. Odintsov,$^{1,2}$\,\thanks{odintsov@ieec.uab.es}
V.K. Oikonomou,$^{3,4}$\,\thanks{v.k.oikonomou1979@gmail.com}}

\affiliation{$^{1)}$ ICREA, Passeig Luis Companys, 23, 08010 Barcelona, Spain\\
$^{2)}$ Institute of Space Sciences (IEEC-CSIC) C. Can Magrans s/n,
08193 Barcelona, Spain\\
$^{3)}$ Laboratory for Theoretical Cosmology, Tomsk State University
of Control Systems
and Radioelectronics, 634050 Tomsk, Russia (TUSUR)\\
$^{4)}$ Tomsk State Pedagogical University, 634061 Tomsk, Russia\\
}

\tolerance=5000

\begin{abstract}
In this work, we introduce a bottom-up $F(R)$ gravity reconstruction
technique, in which we fix the observational indices and we seek for
the $F(R)$ gravity which may realize them. Particularly, as an
exemplification of our method, we shall assume that the
scalar-to-tensor ratio has a specific form, and from it we shall
reconstruct the $F(R)$ gravity that may realize it, focusing on
special values of the parameters in order to obtain analytical
results. The observational indices we study are compatible with the
latest observational data, and we discuss how the functional form of
the observational indices may affect the viability of the model.
\end{abstract}

%PACS numbers: 04.50.Kd, 95.36.+x, 98.80.-k, 98.80.Cq

\maketitle

\section{Introduction}

Describing the inflationary era
\cite{inflation1,inflation2,inflation3,inflation4} in a consistent
way is unarguably one of the streamline tasks of modern cosmology.
Lately, considerable effort is given in describing the early-time
acceleration era by using modified gravity in its various forms
\cite{reviews1,reviews2,reviews4,reviews5}, with $F(R)$ gravity
having a prominent role among all modified gravities. The $F(R)$
gravity framework is a concise and appealing theoretical framework,
which in conjunction with the simplicity, renders $F(R)$ gravity one
of the most important theories of modified gravity. In addition, the
latest Planck data \cite{Ade:2015lrj} and also the BICEP2/Keck-Array
data \cite{Array:2015xqh}, indicate that the $R^2$ inflationary
model is compatible with the observational data, so many researchers
turn their focus on $F(R)$ gravity descriptions of inflation.

Due to the importance of the inflationary era in our primordial
Universe, finding a consistent and observationally acceptable
description is a compelling task. In this line of research, with
this paper we shall approach the $F(R)$ gravity inflationary era by
using a bottom-up approach, in which we shall fix the observational
indices and we shall seek for the vacuum $F(R)$ gravity that can
realize such an evolution, in the slow-roll approximation, with the
observational indices being compatible with the latest observational
data. Particularly, we shall assume that the scalar-to-tensor ratio
$r$ has a specific form, and by using well-known reconstruction
techniques \cite{Nojiri:2009kx}, we shall investigate which $F(R)$
gravity can realize such an evolution, always in the slow-roll
approximation. The slow-roll approximation actually simplifies the
expressions that yield the observational indices in the pure $F(R)$
gravity case. The examples we shall present in order to exemplify
our technique, lead to analytic results, but this is not the general
case. Also our bottom-up approach will enable us to find the exact
form of the Hubble rate as a function of the cosmic time, which
corresponds to the set of the observational indices which we fixed
to have a specific form. Moreover, as we will show, the spectral
index of the primordial curvature perturbations and the
scalar-to-tensor ratio are compatible with the current observational
data. Finally, we briefly discuss various functional forms of the
observational indices and we examine the viability of each scenario,
in the context of slow-roll vacuum $F(R)$ gravity. A different
approach to ours which uses an inverse reconstruction scheme was
presented by  A. Starobinsky in \cite{starobconf}.

This letter is organized as follows: In section II, we present in
brief all the essential information for the vacuum $F(R)$ gravity
theory which are necessary for the calculations that follow. In
section III, we employ well known reconstruction techniques and also
we develop our own technique, which will enable us to find which
$F(R)$ gravity may realize a given set of observational indices
compatible with the observational data. We focus eventually on the
cases that analytical results can be obtained. In section IV we
discuss the viability of various cosmological scenarios for various
functional forms of the scalar-to-tensor ratio. Finally the
conclusions follow in the end of the paper.

In this paper, the cosmological geometric background will be assumed
a flat Friedmann-Robertson-Walker (FRW) geometric background, in
which case the line element is,
\begin{equation}
\label{JGRG14} ds^2 = - dt^2 + a(t)^2 \sum_{i=1,2,3}
\left(dx^i\right)^2\, ,
\end{equation}
where $a(t)$ is the scale factor of the Universe. With regard to the
metric connection, we shall assume that it is a metric compatible,
symmetric and torsion-less connection, the Levi-Civita connection.

\section{Essential Features of $F(R)$ Gravity}

Let us briefly recall some basic features of $F(R)$ gravity, which
are necessary for our presentation, for reviews on this topic see
\cite{reviews1,reviews2,reviews4,reviews5}. The gravitational action
of $F(R)$ gravity in vacuum is equal to,
\begin{equation}\label{action1dse}
\mathcal{S}=\frac{1}{2\kappa^2}\int \mathrm{d}^4x\sqrt{-g}F(R),
\end{equation}
where $\kappa^2$ stands for $\kappa^2=8\pi G=\frac{1}{M_p^2}$ and
also $M_p$ is the Planck mass. By using the metric formalism, we
vary the action with respect to the metric tensor $g_{\mu \nu}$, and
the gravitational equations read,
\begin{equation}\label{eqnmotion}
F'(R)R_{\mu \nu}(g)-\frac{1}{2}F(R)g_{\mu
\nu}-\nabla_{\mu}\nabla_{\nu}F'(R)+g_{\mu \nu}\square F'(R)=0\, ,
\end{equation}
which can be cast as follows,
\begin{align}\label{modifiedeinsteineqns}
R_{\mu \nu}-\frac{1}{2}Rg_{\mu
\nu}=\frac{\kappa^2}{F'(R)}\Big{(}T_{\mu
\nu}+\frac{1}{\kappa^2}\Big{(}\frac{F(R)-RF'(R)}{2}g_{\mu
\nu}+\nabla_{\mu}\nabla_{\nu}F'(R)-g_{\mu \nu}\square
F'(R)\Big{)}\Big{)}\, .
\end{align}
Note that the prime in Eq. (\ref{modifiedeinsteineqns}) denotes
differentiation with respect to the Ricci scalar $R$. For the FRW
metric of Eq. (\ref{JGRG14}), the cosmological equations read,
\begin{align}
\label{JGRG15} 0 =& -\frac{F(R)}{2} + 3\left(H^2 + \dot H\right)
F'(R) - 18 \left( 4H^2 \dot H + H \ddot H\right) F''(R)\, ,\\
\label{Cr4b} 0 =& \frac{F(R)}{2} - \left(\dot H + 3H^2\right)F'(R) +
6 \left( 8H^2 \dot H + 4 {\dot H}^2 + 6 H \ddot H + \dddot H\right)
F''(R) + 36\left( 4H\dot H + \ddot H\right)^2 F'''(R) \, ,
\end{align}
where $H$ denotes the Hubble rate $H=\dot a/a$ and the Ricci scalar
for the FRW metric is equal to $R=12H^2 + 6\dot H$.

\section{Vacuum $F(R)$ Gravity from the Observational Indices}

In this section, by using a bottom-up approach, we shall investigate
how a viable set of the observational indices $n_s$ and $r$ can be
realized by an $F(R)$ gravity in the context of the slow-roll
approximation, where $n_s$ is the power spectrum of the primordial
curvature perturbations and $r$ is the scalar-to-tensor ratio. By
using a well-known reconstruction technique, we shall investigate
which $F(R)$ gravity can realize a given set of observational
indices, starting from the scalar-to-tensor ratio. In order to
illustrate how the method works, we shall use a specific
quantitative example for the scalar-to-tensor ratio. It is important
to note that the slow-roll approximation shall be considered to hold
true during our calculations. In this case, the dynamics of
inflation is quantified perfectly by the generalized slow-roll
indices $\epsilon_1$ ,$\epsilon_2$, $\epsilon_3$, $\epsilon_4$
\cite{Noh:2001ia,Hwang:2001qk,Hwang:2001pu,Kaiser:2010yu,reviews1}.
The first slow-roll parameter $\epsilon_1$ controls the duration of
the inflationary era and more importantly if it occurs in the first
place, and it is equal to $\epsilon_1=-\frac{\dot{H}}{H^2}$. In the
case of vacuum $F(R)$ gravity in the context of the slow-roll
approximation, the slow-roll parameters can be approximated as
follows
\cite{Noh:2001ia,Hwang:2001qk,Hwang:2001pu,Kaiser:2010yu,reviews1},
\begin{equation}
\label{restofparametersfr} \epsilon_2=0\, ,\quad \epsilon_1\simeq
 -\epsilon_3\, ,\quad \epsilon_4\simeq
 \frac{F_{RRR}}{F_{R}}\left( 24\dot{H}+6\frac{\ddot{H}}{H}\right)-3\epsilon_1+\frac{\dot{\epsilon}_1}{H\epsilon_1}\,
 ,
\end{equation}
where $F_R=\frac{\mathrm{d}F}{\mathrm{d}R}$, and
$F_{RRR}=\frac{\mathrm{d}^3F}{\mathrm{d}R^3}$. In addition, the
spectral index of the primordial curvature perturbations of the
vacuum $F(R)$ gravity, and the corresponding scalar-to-tensor ratio,
are equal to
\cite{Noh:2001ia,Hwang:2001qk,Hwang:2001pu,Kaiser:2010yu,reviews1},
\begin{equation}
\label{epsilonall} n_s\simeq 1-6\epsilon_1-2\epsilon_4,\quad
r=48\epsilon_1^2\, .
\end{equation}
We need to stress that the expressions in Eq. (\ref{epsilonall})
hold true only in the context of the slow-roll limit, where
$\epsilon_1,\epsilon_4\ll 1$ and this is an important assumption,
which as we demonstrate it is satisfied.

At this point, let us exemplify our bottom-up reconstruction method
by using a characteristic example, and to this end, let us assume
that the scalar-to-tensor ratio $r$ is equal to,
\begin{equation}\label{model1}
r=\frac{c^2}{(q+N)^2}\, ,
\end{equation}
where $N$ is the $e$-foldings number and $c$, $q$ are arbitrary
parameters for the moment. As we now demonstrate, the choice
(\ref{model1}) can lead to a viable inflationary cosmology. By using
the expression in Eq. (\ref{epsilonall}) for the scalar-to-tensor
ratio $r$, we obtain that,
\begin{equation}\label{diffsqueare}
r=\frac{48 \dot{H}(t)^2}{H(t)^4}
\end{equation}
and by expressing the above expression in terms of the $e$-foldings
number $N$, by using the following,
\begin{equation}\label{trick1}
\frac{\mathrm{d}}{\mathrm{d}t}=H\frac{\mathrm{d}}{\mathrm{d}N}\, ,
\end{equation}
the scalar-to-tensor ratio in terms of $H(N)$ is,
\begin{equation}\label{diffsqueareN}
r=\frac{48 H'(N)^2}{H(N)^2}\, ,
\end{equation}
where the prime now indicates differentiation with respect to $N$.
By combing Eqs. (\ref{model1}) and (\ref{diffsqueareN}), we obtain
the differential equation,
\begin{equation}\label{diffeqns}
\frac{\sqrt{48} H'(N)}{H(N)}=\frac{c}{(q+N)}\, ,
\end{equation}
which can be solved and the solution is,
\begin{equation}\label{hubbleratesol}
H(N)=\gamma  (N+q)^{\frac{c}{4 \sqrt{3}}}\, .
\end{equation}
The spectral index $n_s$ can be calculated in terms of $N$, however
it is worth providing the expression in terms of the cosmic time,
which is,
\begin{equation}\label{nst1}
n_s\simeq 1+\frac{4 \dot{H}(t)}{H(t)^2}-\frac{2 \ddot{H}(t)}{H(t)
\dot{H}(t)}+\frac{F_{RRR}}{F_{R}}\left(
24\dot{H}+6\frac{\ddot{H}}{H}\right)\, ,
\end{equation}
so by using (\ref{hubbleratesol}) and also the following expression,
\begin{equation}\label{asxet1}
\frac{\mathrm{d}^2}{\mathrm{d}t^2}=H^2\frac{\mathrm{d}^2}{\mathrm{d}N^2}+H\frac{\mathrm{d}H}{\mathrm{d}N}\frac{\mathrm{d}}{\mathrm{d}N}\,
,
\end{equation}
the spectral index in terms of the $e$-foldings number is equal to,
\begin{equation}\label{spectralnhnexpr}
n_s\simeq 1+\frac{4 H'(N)}{H(N)}-\frac{2 \left(H(N)
H''(N)+H'(N)^2\right)}{H(N) H'(N)}+ \frac{F_{RRR}}{F_{R}}\Big{(}
24H(N)H'(N)+6H(N)H''(N)+6H'(N)^2\Big{)}\, ,
\end{equation}
where the prime indicates differentiation with respect to the
$e$-foldings number. Finally, by substituting Eq.
(\ref{hubbleratesol}), the spectral index becomes equal to,
\begin{align}\label{spctralindexfinal}
& n_s=1+\frac{c}{\sqrt{3} (N+q)}-\frac{c N}{\sqrt{3}
(N+q)^2}-\frac{c q}{\sqrt{3} (N+q)^2}+\frac{2 N}{(N+q)^2}+\frac{2
q}{(N+q)^2}+\\ \notag & \frac{c^2 \gamma ^2 F_{RRR}
(N+q)^{\frac{c}{2 \sqrt{3}}-2}}{8 F_R}+\frac{5 \sqrt{3} c \gamma ^2
F_{RRR}(N+q)^{\frac{c}{2 \sqrt{3}}-1}}{2 F_R}\, .
\end{align}
As we shall demonstrate later on in this section, the observational
indices (\ref{hubbleratesol}) and (\ref{spctralindexfinal}) can
become compatible with the Planck data \cite{Ade:2015lrj} and even
with the BICEP2/Keck-Array data \cite{Array:2015xqh}, but we need
first to investigate which $F(R)$ gravity can produce the
inflationary era quantified by Eqs. (\ref{hubbleratesol}) and
(\ref{spctralindexfinal}), in order to find the analytic form of the
last two terms in Eq. (\ref{spctralindexfinal}). As we shall see, if
the parameter $c$ is appropriately chosen, an analytic expression
for $F(R)$ can be obtained. In order to find the $F(R)$ gravity
which realizes the observational indices (\ref{hubbleratesol}) and
(\ref{spctralindexfinal}), we shall employ the reconstruction
technique of Ref. \cite{Nojiri:2009kx}, so the cosmological equation
appearing in Eq. (\ref{JGRG15}), can be rewritten in the form,
\begin{equation}\label{frwf1}
-18\left ( 4H(t)^2\dot{H}(t)+H(t)\ddot{H}(t)\right )F_{RR}(R)+3\left
(H^2(t)+\dot{H}(t) \right )F_R(R)-\frac{F(R)}{2}=0\, ,
\end{equation}
where $F'(R)=\frac{\mathrm{d}F(R)}{\mathrm{d}R}$. The reconstruction
method introduced in Ref. \cite{Nojiri:2009kx} uses the $e$-folding
number $N$, which in terms of the scale factor $a$ is,
\begin{equation}\label{efoldpoar}
e^{-N}=\frac{a_0}{a}\, ,
\end{equation}
and in the following we set $a_0=1$. By writing the FRW equation of
Eq. (\ref{frwf1}) in terms of the $e$-foldings number $N$, we
obtain,
\begin{align}\label{newfrw1}
& -18\left ( 4H^3(N)H'(N)+H^2(N)(H')^2+H^3(N)H''(N) \right
)F_{RR}(R)
\\ \notag & +3\left (H^2(N)+H(N)H'(N) \right
)F_R(R)-\frac{F(R)}{2}=0\, ,
\end{align}
where the primes stand for $H'=\mathrm{d}H/\mathrm{d}N$ and
$H''=\mathrm{d}^2H/\mathrm{d}N^2$. By using the function
$G(N)=H^2(N)$, the differential equation (\ref{newfrw1}) can be cast
as follows,
 \begin{align}\label{newfrw1modfrom}
& -9G(N(R))\left ( 4G'(N(R))+G''(N(R)) \right )F_{RR}(R) +\left
(3G(N)+\frac{3}{2}G'(N(R)) \right )F_R(R)-\frac{F(R)}{2}=0\, ,
\end{align}
where $G'(N)=\mathrm{d}G(N)/\mathrm{d}N$ and
$G''(N)=\mathrm{d}^2G(N)/\mathrm{d}N^2$. Also the Ricci scalar can
be expressed in terms of the function $G(N)$ as follows,
\begin{equation}\label{riccinrelat}
R=3G'(N)+12G(N)\, .
\end{equation}
Thus, by solving the differential equation (\ref{newfrw1modfrom}),
we can find the $F(R)$ gravity which may realize a cosmological
evolution. Now we shall make use of the reconstruction technique we
just presented in order to find the $F(R)$ gravity which realizes
the observational indices (\ref{hubbleratesol}) and
(\ref{spctralindexfinal}). In our case, the function $G(N)$ is,
\begin{equation}\label{functiongn}
G(N)=\gamma ^2 (N+q)^{\frac{c}{2 \sqrt{3}}}\, ,
\end{equation}
and consequently, the algebraic equation (\ref{riccinrelat}) takes
the following form,
\begin{equation}\label{rdeterminingr}
12 \gamma ^2 (N+q)^{\frac{c}{2 \sqrt{3}}}+\frac{1}{2} \sqrt{3} c
\gamma ^2 (N+q)^{\frac{c}{2 \sqrt{3}}-1}=R\, .
\end{equation}
In general it is quite difficult to obtain a general solution to
this equation, however is $c$ is chosen appropriately, it is
possible to obtain even full analytic results. For example if
$c=\sqrt{12}$, the results have a fully analytic form. In the
following we shall investigate only the case with $c=\sqrt{12}$, in
which case the algebraic equation (\ref{rdeterminingr}) becomes,
\begin{equation}\label{algebrodiffeqn1}
3 \gamma ^2+12 \gamma ^2 N+12 \gamma ^2 q=R\, ,
\end{equation}
so the function $N(R)$ is equal to,
\begin{equation}\label{nrfunction}
N(R)=\frac{-3 \gamma ^2-12 \gamma ^2 q+R}{12 \gamma ^2}\, .
\end{equation}
By combining Eqs. (\ref{functiongn}) and (\ref{nrfunction})  the
differential equation (\ref{newfrw1modfrom}) in this case becomes,
 \begin{align}\label{newfrw1modfromcaseathand1}
& -36 \gamma ^4 \left(\frac{-3 \gamma ^2-12 \gamma ^2 q+R}{12 \gamma
^2}+q\right)F''(R) +\frac{1}{4} \left(3 \gamma ^2+R\right)
F'(R)-\frac{F(R)}{2}=0\, ,
\end{align}
which can be solved analytically, and the solution is,
\begin{equation}\label{franalytic1}
F(R)=\frac{3}{2} \sqrt{3} \gamma ^3 \delta +\frac{\delta  R^2}{2
\sqrt{3} \gamma }-3 \sqrt{3} \gamma  \delta  R+\mu  \left(R-3 \gamma
^2\right)^{3/2} L_{\frac{1}{2}}^{\frac{3}{2}}\left(\frac{1}{12}
\left(\frac{R}{\gamma ^2}-3\right)\right)\, ,
\end{equation}
where the function $L_n^{\alpha}(x)$ is the generalized Laguerre
Polynomial and also $\delta$ and $\mu$ are arbitrary integration
constants. The existence of the Laguerre polynomial term, imposes
the constraint $R<3\gamma^2$, however in this case the term
containing the root becomes complex. Hence in order to avoid
inconsistencies, we set $\mu=0$, and hence the resulting $F(R)$
gravity is,
\begin{equation}\label{frfinal1}
F(R)=\frac{3}{2} \sqrt{3} \gamma ^3 \delta +\frac{\delta  R^2}{2
\sqrt{3} \gamma }-3 \sqrt{3} \gamma  \delta  R\, ,
\end{equation}
which is a variant form of the Starobinsky model
\cite{Starobinsky:1982ee}. By requiring the coefficient of $R$ to be
equal to one, $\delta$ must be equal to $\delta=-\frac{1}{3 \sqrt{3}
\gamma }$, hence the resulting $F(R)$ gravity during the slow-roll
era is,
\begin{equation}\label{frtermafinal}
F(R)=R-\frac{\gamma ^2}{2}-\frac{R^2}{18 \gamma ^2}\, .
\end{equation}
We can find the Hubble rate as a function of the cosmic time, by
solving the differential equation,
\begin{equation}\label{nthubbleast}
\dot{N}=H(N(t))\, ,
\end{equation}
where $H(N)$ is given in Eq. (\ref{hubbleratesol}), and the
resulting evolution is,
\begin{equation}\label{hubblecaseonecosmict}
H(t)=\frac{1}{4} \left(\Lambda ^2-4 q+\gamma ^2 t^2-2 \gamma \Lambda
t\right)\, ,
\end{equation}
where $\Lambda>0$ is an integration constant. Hence, the resulting
evolution is a quasi-de Sitter evolution, if $\Lambda$ is chosen to
be quite large so that it dominates the evolution at the early-time
era, in which case $H(t)\simeq \frac{\Lambda}{4}$. Also it is
trivial to see that $\ddot{a}>0$, so the solution
(\ref{hubblecaseonecosmict}) describes an inflationary era. Finally,
let us now demonstrate if the resulting cosmology is compatible with
the Planck data. Firstly, let us see how the spectral index becomes
in view of Eq. (\ref{frtermafinal}) and due to the fact that
$F_{RRR}=0$, the spectral index becomes,
\begin{align}\label{spctralindexfinalertermsilan}
& n_s=1+\frac{c}{\sqrt{3} (N+q)}-\frac{c N}{\sqrt{3}
(N+q)^2}-\frac{c q}{\sqrt{3} (N+q)^2}+\frac{2 N}{(N+q)^2}+\frac{2
q}{(N+q)^2}\, .
\end{align}
By using the value of $c$, namely $c=\sqrt{12}$, and also for $N=60$
and $q=-118$, the observational indices become,
\begin{equation}\label{observations1plancbicepresults}
n_s\simeq 0.9658,\,\,\,r\simeq 0.00346842\, .
\end{equation}
Recall that the 2015 Planck data constrain the observational indices
as follows,
\begin{equation}
\label{planckdata} n_s=0.9644\pm 0.0049\, , \quad r<0.10\, ,
\end{equation}
and also, the latest BICEP2/Keck-Array data \cite{Array:2015xqh}
constrain the scalar-to-tensor ratio as follows,
\begin{equation}
\label{scalartotensorbicep2} r<0.07\, ,
\end{equation}
at $95\%$ confidence level. Hence, the observational indices
(\ref{observations1plancbicepresults}) are compatible to both the
Planck and the BICEP2/Keck-Array data.

Hence, by using a bottom-up approach, we found in an analytic way
the $F(R)$ gravity which may realize a viable set of observational
indices $(n_s,r)$. In principle, more choices for the observational
indices are possible, although in most of the cases, semi-analytic
results will be obtained, due to the complexity of the differential
equation (\ref{newfrw1modfrom}).

Another simple case that can be analyzed analytically is the case
for which,
\begin{equation}\label{scalartotensornew}
r=\frac{1}{\beta^2}\, ,
\end{equation}
in which case, by solving the differential equation
(\ref{diffsqueare}), we obtain,
\begin{equation}\label{hubbleratesol23}
H(N)=\gamma  e^{\frac{N}{4 \sqrt{3} \beta }}\, .
\end{equation}
Then, we can easily find that, the function $G(N)$ is,
\begin{equation}\label{functiongn1111}
G(N)=\gamma ^2 e^{\frac{N}{2 \sqrt{3} \beta }}\, ,
\end{equation}
and in effect, the algebraic equation (\ref{riccinrelat}) becomes,
\begin{equation}\label{rdeterminingr1111}
12 \gamma ^2 e^{\frac{N}{2 \sqrt{3} \beta }}+\frac{\sqrt{3} \gamma
^2 e^{\frac{N}{2 \sqrt{3} \beta }}}{2 \beta }=R\, ,
\end{equation}
so the function $N(R)$ is equal to,
\begin{equation}\label{nrfunction1111}
N(R)=2 \sqrt{3} \beta  \ln \left(\frac{2 \beta  R}{\left(24 \beta
+\sqrt{3}\right) \gamma ^2}\right)\, .
\end{equation}
By combining Eqs. (\ref{functiongn1111}) and (\ref{nrfunction1111})
the differential equation (\ref{newfrw1modfrom}) in the case at hand
becomes,
 \begin{align}\label{newfrw1modfromcaseathand11111}
& -36 \gamma ^4 \left(-\frac{3 \left(8 \sqrt{3} \beta +1\right)
R^2}{\left(24 \beta +\sqrt{3}\right)^2}\right)F''(R) +\frac{\left(12
\beta +\sqrt{3}\right) R}{2 \left(24 \beta +\sqrt{3}\right)}
F'(R)-\frac{F(R)}{2}=0\, ,
\end{align}
which can also be solved analytically, and the resulting $F(R)$
gravity is,
\begin{equation}\label{franalytic11111}
F(R)=C_1 R^{\mu}+C_2R^{\nu}\, ,
\end{equation}
where $C_1$ and $C_2$ are integration constants and also $\mu$ and
$\nu$ stand for,
\begin{align}\label{muandnu}
& \mu=\frac{96 \beta ^2+\frac{\sqrt{24 \beta +\sqrt{3}} \sqrt{384
\sqrt{3} \beta ^3-912 \beta ^2-32 \sqrt{3} \beta
+1}}{\sqrt[4]{3}}+28 \sqrt{3} \beta +3}{32 \sqrt{3} \beta +4},\\
\notag & \nu=\frac{96 \beta ^2-\frac{\sqrt{24 \beta +\sqrt{3}}
\sqrt{384 \sqrt{3} \beta ^3-912 \beta ^2-32 \sqrt{3} \beta
+1}}{\sqrt[4]{3}}+28 \sqrt{3} \beta +3}{32 \sqrt{3} \beta +4}\, .
\end{align}
Some aspects of the $F(R)$ gravity (\ref{franalytic11111}) were
investigated in Ref. \cite{Sebastiani:2013eqa}. By combining Eqs.
(\ref{nrfunction1111}) and (\ref{franalytic11111}), the term
$\frac{F_{RRR}}{F_R}$ can easily be calculated, so the resulting
expression for the spectral index $n_s$ is,
\begin{align}\label{nsfinalmodel2}
& n_s=1+\frac{4 \beta ^2 C_1 2^{\nu } e^{\frac{N}{\sqrt{3} \beta }}
\left(\frac{\gamma ^2 e^{\frac{N}{2 \sqrt{3} \beta }}}{8 \beta
^2}+\frac{5 \sqrt{3} \gamma ^2 e^{\frac{N}{2 \sqrt{3} \beta }}}{2
\beta }\right) \left((\mu -2) (\mu -1) \mu  \left(\left(24 \beta
-\sqrt{3}\right)^{\mu } \gamma ^{2 \mu } e^{-\frac{\mu  N}{2
\sqrt{3} \beta }}\right)\right)}{\beta ^{\mu }
\left(\left(\sqrt{3}-24 \beta \right)^2 \gamma ^4 \left(\frac{C_1
\mu  2^{\nu } \left(\left(24 \beta -\sqrt{3}\right)^{\mu } \gamma
^{2 \mu } e^{-\frac{\mu  N}{2 \sqrt{3} \beta }}\right)}{\beta ^{\mu
}}+\frac{C_2 2^{\mu } \nu  \left(\left(24 \beta
-\sqrt{3}\right)^{\nu } \gamma ^{2 \nu } e^{-\frac{\nu  N}{2
\sqrt{3} \beta }}\right)}{\beta ^{\nu }}\right)\right)}\\ \notag &
+\frac{4 \beta ^2 C_2 2^{\mu } e^{\frac{N}{\sqrt{3} \beta }}
\left(\frac{\gamma ^2 e^{\frac{N}{2 \sqrt{3} \beta }}}{8 \beta
^2}+\frac{5 \sqrt{3} \gamma ^2 e^{\frac{N}{2 \sqrt{3} \beta }}}{2
\beta }\right) \left((\nu -2) (\nu -1) \nu  \left(\left(24 \beta
-\sqrt{3}\right)^{\nu } \gamma ^{2 \nu } e^{-\frac{\nu  N}{2
\sqrt{3} \beta }}\right)\right)}{\beta ^{\nu }
\left(\left(\sqrt{3}-24 \beta \right)^2 \gamma ^4 \left(\frac{C_1
\mu  2^{\nu } \left(\left(24 \beta -\sqrt{3}\right)^{\mu } \gamma
^{2 \mu } e^{-\frac{\mu  N}{2 \sqrt{3} \beta }}\right)}{\beta ^{\mu
}}+\frac{C_2 2^{\mu } \nu  \left(\left(24 \beta
-\sqrt{3}\right)^{\nu } \gamma ^{2 \nu } e^{-\frac{\nu  N}{2
\sqrt{3} \beta }}\right)}{\beta ^{\nu }}\right)\right)}\, .
\end{align}
The parameter space contains a lot of free parameters, and
specifically $\beta$, $\gamma$ and the integration constants $C_1$
and $C_2$, hence the compatibility with the Planck and
BICEP2/Keck-array data of Eqs. (\ref{planckdata}) and
(\ref{scalartotensorbicep2}), can easily be achieved. Indeed, if for
example we choose $\beta=5.6201$, for $N=60$ $e$-foldings and with
the rest of the parameters being equal to one, the spectral index
becomes $n_s\simeq 0.966983$ and the scalar-to-tensor ratio becomes
equal to $r=0.0316601$, so compatibility with both the Planck
(\ref{planckdata}) and the BICEP2/Keck-Array data
(\ref{scalartotensorbicep2}) is achieved.

\section{General Forms of the Observational Indices: A Critical Discussion}

An interesting outcome may be obtained if we study the various forms
that the observational indices may take. In principle, any
combination of functions is allowed, however in this section we
shall consider exponentials and logarithmic functions that may
appear in the scalar-to-tensor ratio, and we shall study the
implications of these functional forms of the scalar-to-tensor
ratio. In principle, the resulting cosmology should be critically
checked in order to validate that it describes an inflationary
cosmology, exactly as we did in the previous section. However, we
shall attempt a superficial approach focusing only on the
observational indices, to see their behavior when the aforementioned
functions are used. Also we shall take into account the simplest
functional forms of the observational indices, and we shall assume
that the theory is a slow-roll $F(R)$ gravity. Let us start with the
exponential case, and we assume that the scalar-to-tensor ratio has
the following form,
\begin{equation}\label{scalartotensorratioexp}
r=\frac{\gamma}{e^{\beta N}}\, ,
\end{equation}
and hence, since the scalar-to-tensor ratio in terms of $H(N)$ is
given in Eq. (\ref{diffsqueareN}), by combining Eqs.
(\ref{diffsqueareN}) and (\ref{scalartotensorratioexp}), we obtain
the following differential equation,
\begin{equation}\label{diffeqnexp}
\frac{\sqrt{48} H'(N)}{H(N)}=\frac{\sqrt{\gamma }}{\exp
\left(\frac{\beta  N}{2}\right)}\, ,
\end{equation}
which can be solved and the resulting Hubble rate is,
\begin{equation}\label{expnentsolq}
H(N)=c \,e^{-\frac{\sqrt{\gamma } e^{-\frac{1}{2} (\beta  N)}}{2
\sqrt{3} \beta }}\, ,
\end{equation}
where $c$ is an integration constant. By substituting Eq.
(\ref{expnentsolq}) in Eq. (\ref{spectralnhnexpr}), the spectral
index $n_s$ reads,
\begin{equation}\label{spectralindexexp}
n_s\simeq 1+\beta +\frac{\gamma  c^2 F_{RRR} e^{\beta
(-N)-\frac{\sqrt{\gamma } e^{-\frac{1}{2} (\beta  N)}}{\sqrt{3}
\beta }}}{8 F_R}+\frac{5 \sqrt{3} \sqrt{\gamma } c^2 F_{RRR}
e^{-\frac{\sqrt{\gamma } e^{-\frac{1}{2} (\beta N)}}{\sqrt{3} \beta
}-\frac{\beta  N}{2}}}{2 F_R}\, .
\end{equation}
It is easy to understand how the spectral index above behaves, even
without knowing the exact form of the $F(R)$ gravity, since the
presence of the exponentials renders the last two terms negligible,
and hence the spectral index reads $n_s\simeq 1+\beta$. The last
expression for the spectral index cannot be compatible with the
current observational data for any value of $\beta>0$, hence vacuum
$F(R)$ gravity theories which yield a scalar-to-tensor ratio of the
form (\ref{scalartotensorratioexp}) do not yield a viable
inflationary era.

Another class of models which we consider is the ones for which the
scalar-to-tensor ratio has the following form,

\begin{equation}\label{scalartotensorratioexp1}
r=\frac{\gamma}{\ln N}\, ,
\end{equation}
and hence  by combining Eqs. (\ref{diffsqueareN}) and
(\ref{scalartotensorratioexp1}), we obtain the following
differential equation,
\begin{equation}\label{diffeqnexp1}
\frac{\sqrt{48} H'(N)}{H(N)}=\frac{\sqrt{\gamma }}{\sqrt{\ln N}}\, ,
\end{equation}
which can be solved analytically and the resulting Hubble rate is,
\begin{equation}\label{expnentsolq1}
H(N)=c\, e^{\frac{1}{2} \sqrt{\frac{\pi }{3}} \sqrt{\gamma }
\text{Erfi}\left(\sqrt{\log \left(\frac{N}{2}\right)}\right)}\, ,
\end{equation}
where $c$ is again an integration constant and the function
$\mathrm{Erfi}(x)$ is the imaginary error function. By substituting
Eq. (\ref{expnentsolq1}) in Eq. (\ref{spectralnhnexpr}), the
spectral index $n_s$ reads in this case,
\begin{equation}\label{spectralindexexp1}
n_s\simeq 1+\frac{5 \sqrt{3} \sqrt{\gamma } c^2
F_{RRR}e^{\sqrt{\frac{\pi }{3}} \sqrt{\gamma }
\text{Erfi}\left(\sqrt{\ln \left(\frac{N}{2}\right)}\right)}}{2 F_R
\sqrt{\ln \left(\frac{N}{2}\right)}}+\frac{\gamma  c^2 F_{RRR}
e^{\sqrt{\frac{\pi }{3}} \sqrt{\gamma } \text{Erfi}\left(\sqrt{\ln
\left(\frac{N}{2}\right)}\right)}}{8 F_R \ln
\left(\frac{N}{2}\right)}+\frac{1}{N \ln \left(\frac{N}{2}\right)}\,
.
\end{equation}
In this case, if the parameter $c$ is appropriately chosen, and if
the resulting $F(R)$ gravity yields the term
$\frac{F_{RRR}}{F_{R}}<0$, then the spectral index may be compatible
with the observational data. However, it is not easy to find the
exact form of the $F(R)$ gravity, since by applying the
reconstruction method of the previous section, one may see that the
resulting differential equations cannot be solved analytically,
unless an approximation is used.

Hence in this section we showed that apart from the power-law
functions of the $e$-foldings number, other functions may
successfully describe a viable $F(R)$ gravity inflation, but
analyticity is hard to achieve. Also, combinations of the
exponential and logarithmic functions, with power law functions may
also yield quite interesting results. One such application occurs
for the constant-roll inflationary scenario, but we defer this to a
future work.

\section{Conclusions}

In this letter we studied an $F(R)$ reconstruction technique by
using a bottom-up approach, in which the observational indices are
fixed. Particularly, we assumed a specific form for the
scalar-to-tensor ratio, in the context of slow-roll $F(R)$ gravity,
and we investigated from which $F(R)$ gravity this scalar-to-tensor
ratio may be realized. Also we calculated the spectral index of the
primordial curvature perturbations, if the scalar-to-tensor ratio
has the assumed specific form. By using well-known reconstruction
techniques, we were able to find the analytic form of the $F(R)$
gravity and in addition, the observational indices can be compatible
with the observational data. We also studied several functional
forms of the scalar-to-tensor ratio, such as exponential and
logarithmic functions, and we discussed the behavior of the spectral
index and in effect the viability of the theory.

In principle, the bottom-up approach we propose with this letter,
can be applied to more complex functional dependencies of the
scalar-to-tensor ratio, but it is compelling in the end to find the
Hubble rate as a function of the cosmic time, and check explicitly
whether the resulting evolution is an inflationary evolution.
Finally, apart from the slow-roll reconstruction method we studied
in this paper, one can also study the constant-roll generalization
of our bottom-up reconstruction technique. This issue will be
addressed in future work.

\section*{Acknowledgments}

This work is supported by MINECO (Spain), FIS2016-76363-P and by
CSIC I-LINK1019 Project (S.D.O).

\end{document}